\documentclass[a4paper]{PoS}
\usepackage{multirow}
\title{Colour Reconnection - Models and Tests}

\usepackage{floatrow}
\newfloatcommand{capbtabbox}{table}[][\FBwidth]

\ShortTitle{Colour Reconnection - Models and Tests}

\author{\speaker{Jesper Roy Christiansen}\footnote{Work is supported in part by the MCnetITN FP7 Marie Curie Initial
Training Network, contract PITN-GA-2012-315877 and the Swedish Research
Council, contract 621-2013-4287.}\\
  Department of Astronomy and Theoretical Physics, Lund University, S\"olvegatan 14A, SE-223 62 Lund, Sweden\\
        E-mail: \email{jesper.christiansen@thep.lu.se}}

\abstract{Recent progress on colour reconnection within the \textsc{Pythia}
  framework is presented. A new model is introduced, based on the SU(3)
  structure of QCD and a minimization of the potential string energy. The
  inclusion of the epsilon structure of SU(3) gives a new baryon production
  mechanism and makes it possible simultaneously to describe hyperon
  production at both e$^+$e$^-$ and pp colliders. Finally, predictions for
  e$^+$e$^-$ colliders, both past and potential future ones, are presented. \\
\\ \flushright{LU-TP 15-45 \\ MCNET-15-29} \\}

\FullConference{The European Physical Society Conference on High Energy Physics\\
		22--29 July 2015\\
		Vienna, Austria}
\begin{document}

\section{Introduction}
This conference proceeding is based on work carried out by the author in collaboration with Peter Skands~\cite{Christiansen:2015yqa}, Torbj{\"o}rn Sj{\"o}strand~\cite{Christiansen:2015yca} and Christian Bierlich~\cite{Bierlich:2015rha}, respectively. The respective publications contain more information on each individual topic, and this is mainly intended as a short summary.

The start of the LHC sparked a renewed interest in Colour Reconnection (CR). This is mainly due to two observations: firstly, the $\Lambda$ production rate is observed to be significantly above the expected one~\cite{Aamodt:2011zza,Khachatryan:2011tm}. Secondly, flow-like effects are observed not only in heavy ion collisions but also in pp collisions~\cite{Ortiz:2013yxa}. Several new models have then been developed~\cite{Christiansen:2015yqa,Bierlich:2014xba,Pierog:2013ria}; all capable of describing the $\Lambda$ production. This paper will predominantly focus on a new CR model implemented in \textsc{Pythia}~8~\cite{Sjostrand:2014zea}. The main extension in the new model is the inclusion of junction structures, which naturally leads to a baryon enhancement, which can explain the $\Lambda$ enhancement.

Another important point is not only to suggest new models, but also to consider good observables to test these models. Due to the postulated jet universality between e$^+$e$^-$ and pp colliders, the $\Lambda$ enhancement should be due to physics at pp colliders which can be neglected at e$^+$e$^-$ colliders. The key observation that most models rely on is the increased final state activity present at pp colliders, which then alters the hadronization (either through CR or by directly changing hadronization probabilities). An obvious observable is therefore to consider identified hadron production as a function of the final state activity, which can be related to the charged multiplicity. Exactly this will be presented for several models in section \ref{sec:pp}. 

The major interest has lately been on pp colliders for natural reasons, but CR also affects e$^+$e$^-$ colliders. Even though the effects are much smaller, the cleaner environment may allow potential large constraints on new CR models. It will at least provide another handle to probe CR effects. CR was already studied at LEP, but no final conclusions could be drawn due to limited statistics. With a new e$^+$e$^-$ collider in the future, it will be possible to get significantly more statistics and thereby also a much better handle on CR. It also opens up for observables not considered at LEP, for instance the mass measurement of the W boson in the fully hadronic channel can then be used a measurement of CR. The uncertainty from CR should also be included as soon as the final state contains hadrons, especially given the high expected precision.

The paper is structured as follows: in section \ref{sec:newModel} a short overview of the CR model is presented. This is followed by first results at pp colliders (section \ref{sec:pp}) and then e$^+$e$^-$ colliders (section \ref{sec:ee}). Finally, a summary and outlook is presented in section \ref{sec:conclusion}.

\section{The new CR model} \label{sec:newModel}
\begin{figure}
  \centering
  \includegraphics[width=0.8\textwidth]{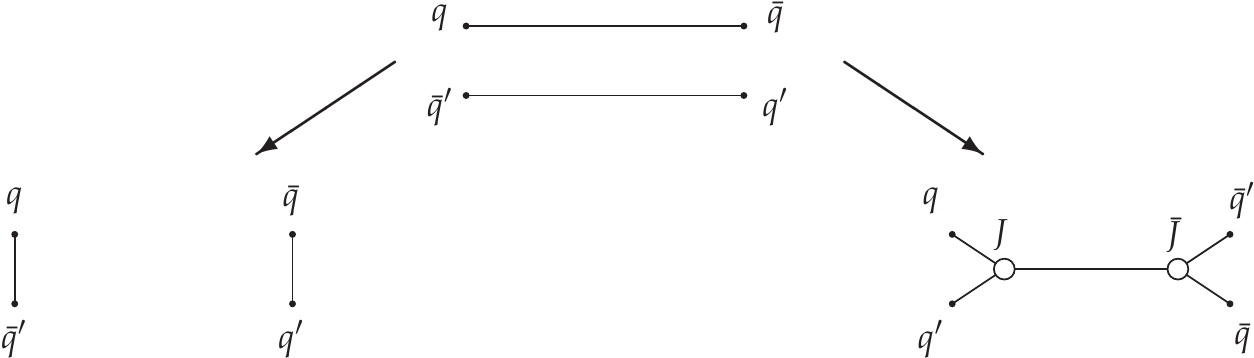}
  \caption{ \label{fig:colrec} Sketch of how two $q\bar{q}$ dipoles (top) can be
    reconnected to different colour topologies (left and right). The right
    connection gives rise to a double 
    junction, which in turn will produce baryons. Notice that the placement of
    the pairs differs in the junction figure.} 
\end{figure}

The new CR model in \textsc{Pythia} is applied just prior to the
hadronization. It takes the leading-colour ($N_c \rightarrow \infty$) strings 
and transform them to a different colour configuration based on three
principles: firstly the 
SU(3) colour rules from QCD determine if two strings are colour compatible
(e.g. there is only a $1/9$ probability that the top configuration of
fig.~\ref{fig:colrec} can transform to the left configuration purely from
colour considerations). Secondly a
simplistic space-time picture to check causal contact between the
strings. Finally the $\lambda$ measure (which is a string-length measure) to
decide 
whether a possible reconnection is 
actually favoured. Since the model relies purely on the outgoing partons, it is applicable to any type of collision. The main extension
compared to the other CR models in \textsc{Pythia} is the introduction of
reconnections that form junction structures (fig.~\ref{fig:colrec}). From a pure colour consideration the
probability to form a junction topology is three times larger than an
ordinary reconnection. The
junction will introduce additional strings, however, and it is therefore often
disfavoured due to a larger $\lambda$ measure. Given the close connection
between junctions and baryons, the new model predicts a baryon enhancement.
It was shown to be able to simultaneously describe the $\Lambda$
production for both LEP and LHC experiments, which neither of the earlier
\textsc{Pythia} tunes have been able to do. 

\section{Comparison to pp data} \label{sec:pp}
\begin{figure}
  \centering
  \includegraphics[scale=0.5]{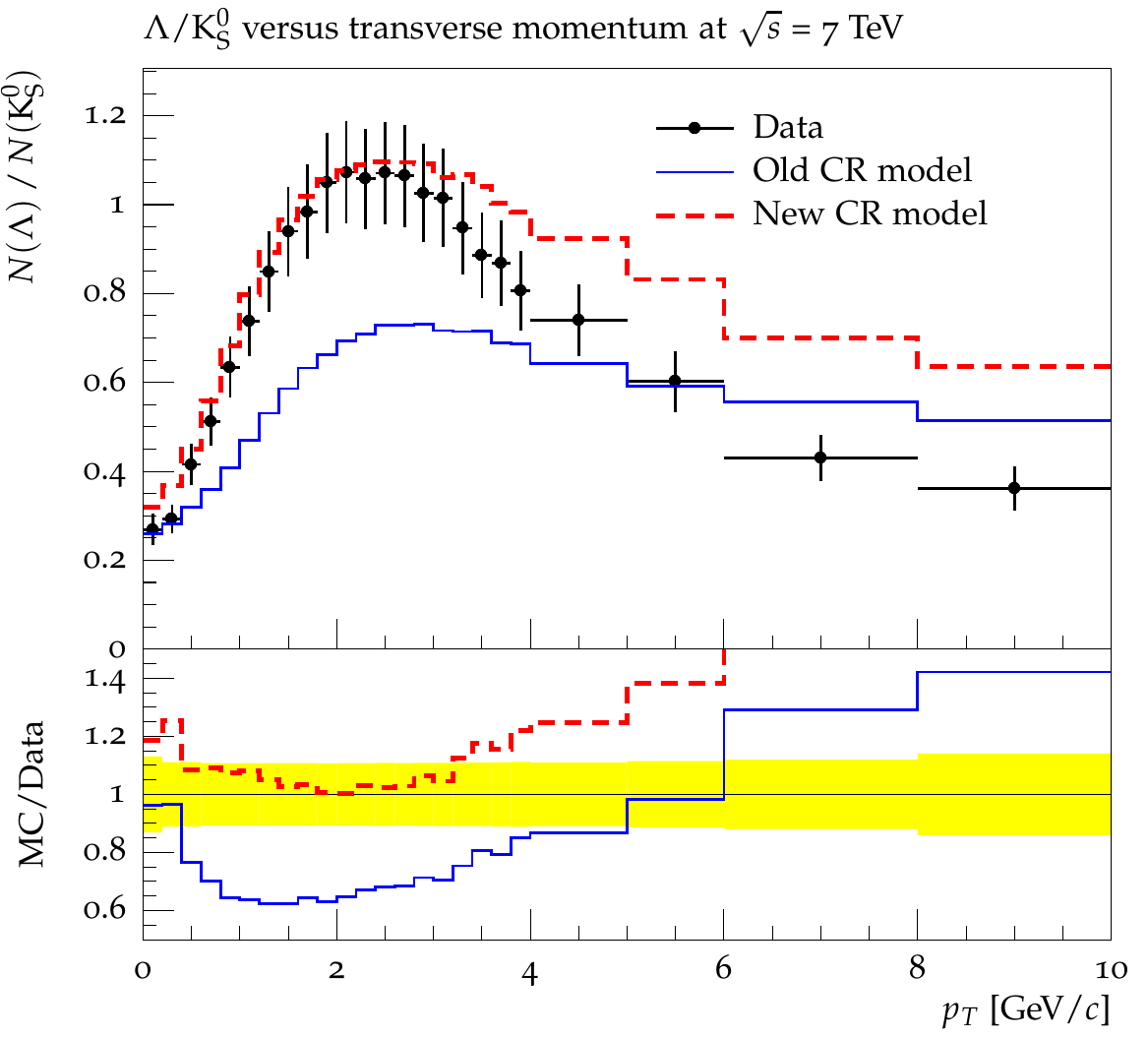}
  \caption{\label{fig:lambdapT}
    The $\Lambda / \mathrm{K}_\mathrm{s}^0$ $p_\perp$-distribution as
    measured by the CMS experiment~\cite{Khachatryan:2011tm}. All \textsc{Pythia} 
    simulations were NSD with a lifetime cut-off ($\tau_\mathrm{max} = 10$
    mm/c) and a rapidity cut on 2 ($|y| < 2$). The yellow error band represents the
    experimental $1\sigma$ deviation.}
\end{figure}

The natural first observable to consider is $\Lambda / \mathrm{K}_\mathrm{s}^0$ production (fig. \ref{fig:lambdapT}). The new model does a significantly better job at describing this observable, especially in the mid-$p_\perp$ region between $\sim$~$1-3$~GeV. It should be recalled that the new model is tuned to get the overall amount of $\Lambda$ particles correct. The high-$p_\perp$ tail is still not well described and is an area which would be interesting to study in more detail. However, the majority of all the $\Lambda$ particles are produced well below this region.

\begin{figure}
\centering
\includegraphics[width=0.7\textwidth]{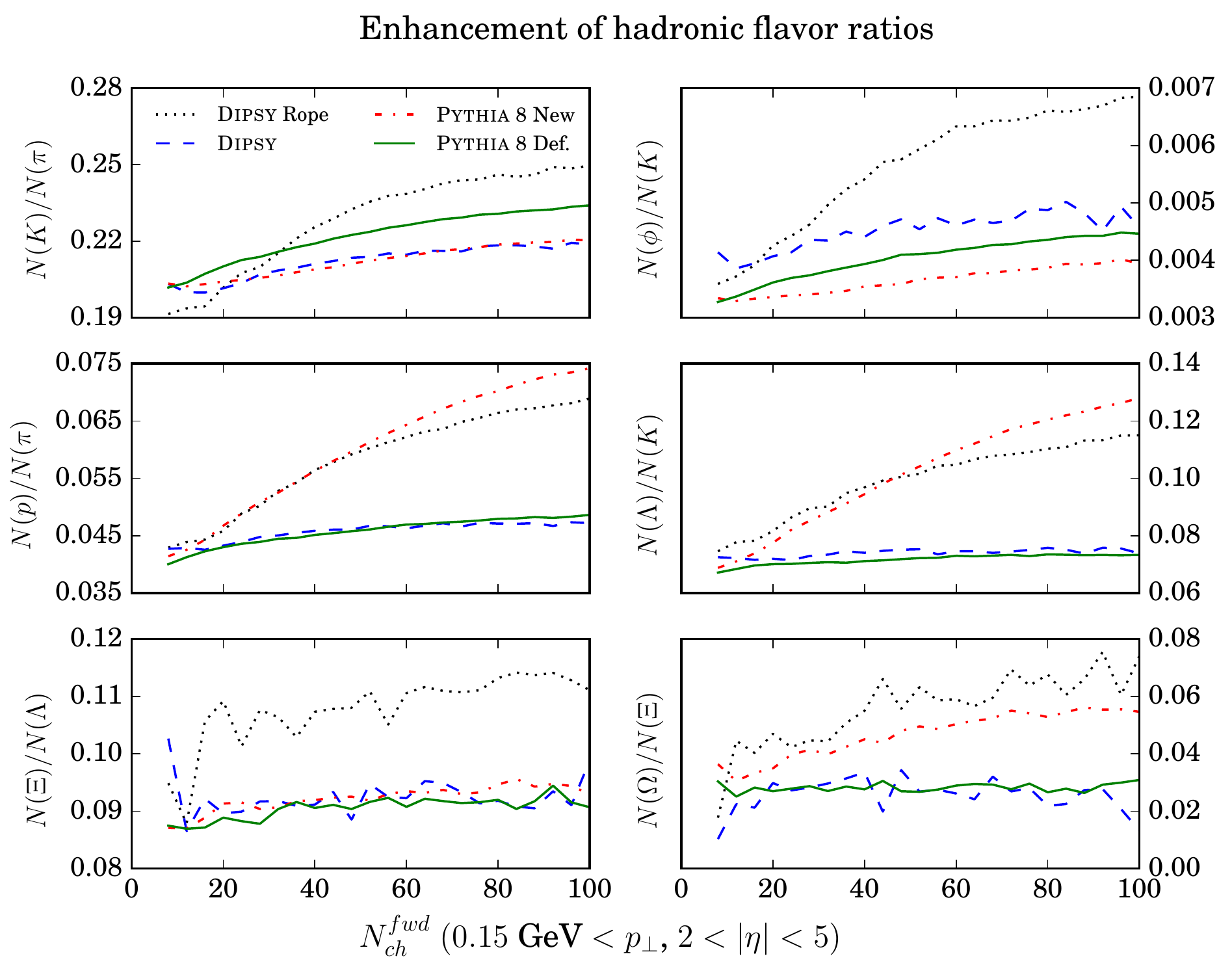}
\caption{\label{fig:ratios}Ratios of identified hadrons as functions of $N_{ch}^{fwd}$ at
  $\sqrt{s} = 13$ TeV. The top row shows meson ratios with the numerator
  having one more strange quark than the denominator. The middle row shows baryon to meson
  ratios, with same amount of strange quarks. The bottom row shows baryon
  ratios with the numerator
  having one more strange quark than the denominator. Note that the vertical
  axis differs between the figures and that zero is suppressed.}
\end{figure}

Ratios between different identified particles yields as a function of charged multiplicity are shown in fig.~\ref{fig:ratios}. As was already hinted in the introduction, these are a great observables to test the new CR models. In addition to the model described, this figure also contains the DIPSY model. The rope extension to the DIPSY model both contains a baryon and a strangeness enhancement. Both the $\Lambda /K$ and $p/\pi$ ratios show a clear increase with multiplicity for the new models. This is exactly what was expected, due to the baryon enhancement. One thing to note is that the baseline is extremely flat, thereby providing an excellent probe for new models. The new CR model in \textsc{Pythia} does not contain any strangeness enhancement, which can be seen for the other distributions where no real enhancements are seen. The only exception is the $\Omega$ production. This is due to a very strong suppression in the ordinary production channel, which is not present for production through junctions. The DIPSY rope model shows an enhancement for all observables, due to the strangeness enhancement. The different predictions for strangeness can used to tell which model best describe the data. It should be noted that these observables are not only good to distinguish between exactly these two models, but can provide clear information about both strangeness and baryon enhancement regardless of the model.

One of the classic observables for radial flow used in heavy ion collisions is to study the $p_\perp$ spectrum of $\Lambda / K$ for different centralities. For smaller centralities, larger multiplicities, the flow effects are larger, thereby pushing the ``peak'' to larger $p_\perp$ values. A similar measurement can be done in pp, but using multiplicity as the centrality measure (fig.~\ref{fig:ratios}). The new CR model in \textsc{Pythia} shows exactly the same qualitative behaviour. Again this highlight the connection between flow and CR, which is an area, where more studies would be of great interest. 

\begin{figure}[!bth]
\centering
\includegraphics[width=0.7\textwidth]{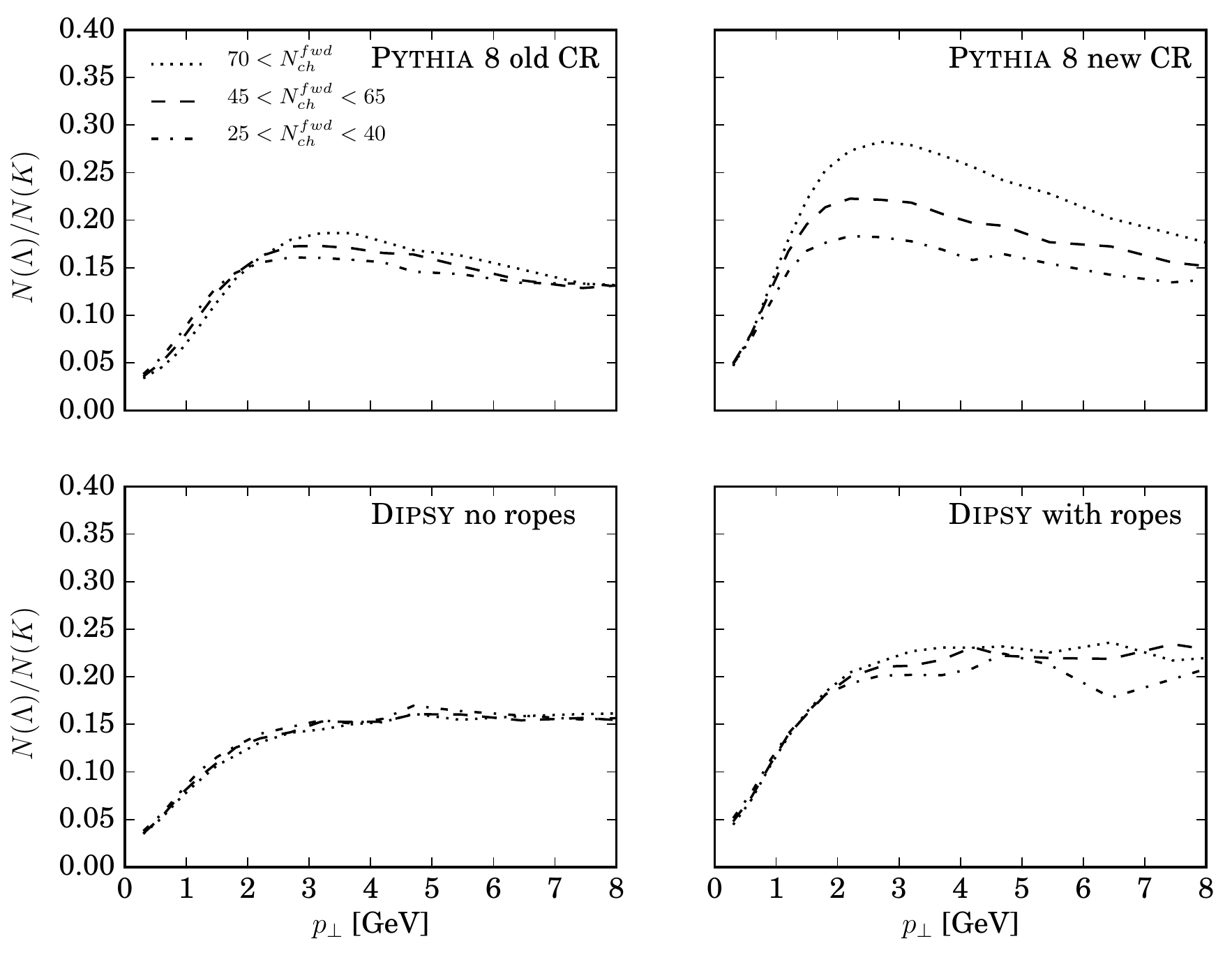}
\caption{\label{fig:flowlike} Ratio of $\Lambda/K$ as a function of $p_\perp$ in three bins of $N_{ch}^{fwd}$. In the right column the new colour reconnection models are shown, and in the left column the old ones.}
\end{figure}

\section{Comparison to e$^+$e$^-$ data}\label{sec:ee}
\begin{figure}
\begin{floatrow}
\capbtabbox{%
  \begin{tabular}{|c|c|c|c|}
    \hline
    \multirow{2}{*}{Model} & \multicolumn{3}{|c|}{$\langle \delta                                                                                                                                       
      \overline{m}_{\mathrm{W}} \rangle$ (MeV)}\\
    \cline{2-4}
    & 170 GeV & 240 GeV & 350 GeV \\
    \hline
    SK-I       & +18 & +95  & +72  \\
    SK-II      & -14 & +29  & +18  \\
    SK-II'     & -6  & +25  & +16  \\
    GM-I    & -41 & -74  & -50  \\
    GM-II   & +49 & +400 & +369 \\
    GM-III  & +2  & +104 & +60  \\
    CS      & +7  & +9   & +4   \\
    \hline
  \end{tabular}

}{%
\caption{\label{tab:wMass}Systematic $\mathrm{W}$ mass shifts at three different center-of-mass energies. The SK and GM models are different CR models, and CS is the new CR model presented in this study. The shifts shown are between no CR and the CR model stated.}%
}

\ffigbox{%
  \centering
  \includegraphics*[width=0.5\textwidth]{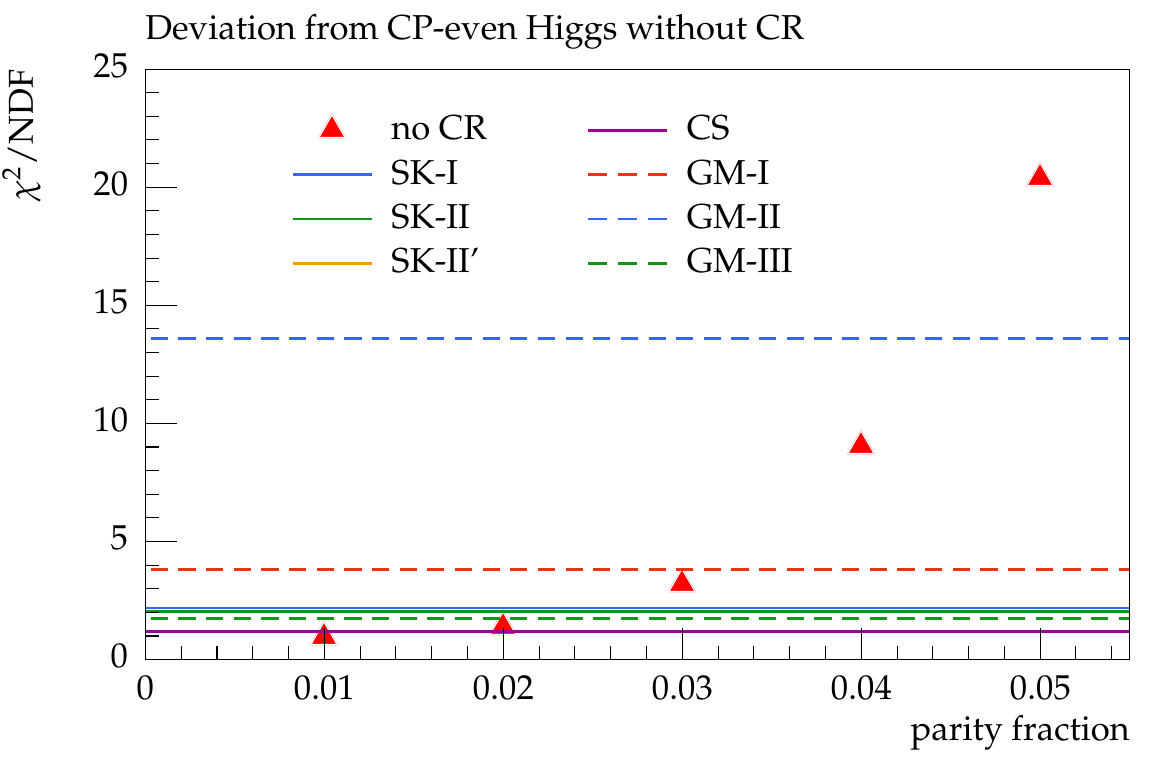}
}{%
  \caption{Deviations between a $CP$-even Higgs without CR and models with either
  increased $CP$-oddness or a CR model. The deviation is quantified as the
  $\chi^2/\mathrm{NDF}$ deviation for the angle between the W, containing the leading jet, and its decay product ($\theta_1$).
\label{ar3:fig:chi2}}%
}
\end{floatrow}
\end{figure}

It has already been well established that CR shifts the mass measurement of the W boson in the fully hadronic channel. This gave a large systematic uncertainty at LEP. This study turns the tables around and use the shift as a probe of CR. The expected statistical precision for the W mass measurement is below 1~MeV, clearly enough to distinguish between the different CR scenarios (tab. \ref{tab:wMass}). Another intriguing observation is the large center-of-mass dependency of the shifts. Two competing effects are at play here: larger energies means lower probability for a CR to occur, but a single reconnection has a larger effect. The effects are seen to be largest at intermediate energies.

As is clear from the W mass study above, CR will affect observables at e$^+$e$^-$ colliders. It therefore has to be included as an uncertainty for other observables. One example is the Higgs parity measurement in a fully hadronic decay (i.e. H$^0 \rightarrow$W$^+$W$^- \rightarrow \mathrm{q}\bar{\mathrm{q}}\mathrm{q}\bar{\mathrm{q}}$). This measurement relies heavily on the angles between the observed jets, which are known to be sensitive to CR. To study the size of the effects a simple $\chi^2$ comparison was carried out. Different mixtures of a CP-even and CP-odd Higgs were considered together with different CR models for the fully CP-even Higgs (fig.~\ref{ar3:fig:chi2}). The effects of CR are seen to be of the order of a few percent, and thus if higher precision is statistically possible, CR needs to be included as an uncertainty. This should not necessarily be seen as a lower limit on the obtainable precision, but rather as the point when CR needs to be considered.

%


%

\section{Summary and outlook} \label{sec:conclusion}
The implementation of a new CR model in \textsc{Pythia} is described and its effects at both pp and e$^+$e$^-$ colliders are considered.
The new model correctly describe the $\Lambda$ production at pp colliders, due to the baryon enhancement from the junction production mechanism. 

It is important to note that this should not be seen as the final step in our understanding of CR, but merely as a step along the way. The new model still has problems describing observables (e.g. $\langle p_\perp \rangle$ vs. mass), and there is thus room for improvement in the model building. Also the experimental data used to constrain these models can be enhanced by including additional observables, for instance some of the observables suggested in this paper. Similarly for e$^+$e$^-$ colliders, the story is not finished and additional more detailed studies, both experimental and phenomenological, will be needed.


\begin{thebibliography}{99}

\bibitem{Christiansen:2015yqa}
  J.~R.~Christiansen and P.~Z.~Skands,
  JHEP {\bf 1508} (2015) 003
  [arXiv:1505.01681 [hep-ph]].

\bibitem{Christiansen:2015yca}
  J.~R.~Christiansen and T.~Sj{\"o}strand,
  Eur.\ Phys.\ J.\ C {\bf 75} (2015) 9,  441
  [arXiv:1506.09085 [hep-ph]].

\bibitem{Bierlich:2015rha} 
  C.~Bierlich and J.~R.~Christiansen,
  [arXiv:1507.02091 [hep-ph]].

\bibitem{Aamodt:2011zza}
  K.~Aamodt {\it et al.} [ALICE Collaboration],
  Eur.\ Phys.\ J.\ C {\bf 71} (2011) 1594
  [arXiv:1012.3257 [hep-ex]].

\bibitem{Khachatryan:2011tm}
  V.~Khachatryan {\it et al.} [CMS Collaboration],
  JHEP {\bf 1105} (2011) 064
  [arXiv:1102.4282 [hep-ex]].

\bibitem{Ortiz:2013yxa}
  A.~Ortiz Velasquez {\it et al.},
  Phys.\ Rev.\ Lett.\  {\bf 111} (2013) 4,  042001
  [arXiv:1303.6326 [hep-ph]].

\bibitem{Bierlich:2014xba}
  C.~Bierlich {\it et al.},
  JHEP {\bf 1503} (2015) 148
  [arXiv:1412.6259 [hep-ph]].

\bibitem{Pierog:2013ria}
  T.~Pierog {\it et al.},
  Phys.\ Rev.\ C {\bf 92} (2015) 3,  034906
  [arXiv:1306.0121 [hep-ph]].

\bibitem{Sjostrand:2014zea}
  T.~Sj{\"o}strand {\it et al.},
  Comput.\ Phys.\ Commun.\  {\bf 191} (2015) 159
  [arXiv:1410.3012 [hep-ph]].



\end{thebibliography}
\end{document}